\documentclass[12pt]{iopart}
\usepackage{graphicx,color}
\usepackage{rotating}
\begin{document}
\title[Vibrational excitons in ionophores]{Vibrational excitons in ionophores; Experimental probes for quantum coherence-assisted ion transport and selectivity in ion channels}
\author{Ziad Ganim$^1$, Andrei Tokmakoff$^2$ and Alipasha Vaziri$^{3,4}$}
\address{$^1$ Physik-Department E22, Technische Universit{\"a}t M{\"u}nchen, James-Franck-Str. 1, D-85748 Garching, Germany}
\address{$^2$ Department of Chemistry, Massachusetts Institute of Technology, 77 Massachusetts Ave., Cambridge, MA 02139, USA}
\address{$^4$ Max F. Perutz Laboratories (MFPL), Universit{\"a}t Wien, Dr. Bohr-Gasse 9, A-1030 Wien, Austria}
\address{$^3$ Research Institute of Molecular Pathology (IMP), Dr. Bohr-Gasse 7, A-1030 Wien, Austria}
\ead{alipasha.vaziri@univie.ac.at}
\begin{abstract}
Despite a large body of work, the exact molecular details underlying ion-selectivity and transport in the potassium channel have not been fully laid to rest. One major reason has been the lack of experimental methods that can probe these mechanisms dynamically on their biologically relevant time scales. Recently it was suggested that quantum coherence and its interplay with thermal vibration might be involved in mediating ion-selectivity and transport. In this work we present an experimental strategy for using time resolved infrared spectroscopy to investigate these effects. We show the feasibility by demonstrating the IR absorption and Raman spectroscopic signatures of potassium binding model molecules that mimic the transient interactions of potassium with binding sites of the selectivity filter during ion conduction. In addition to guide our experiments on the real system we have performed molecular dynamic-based simulations of the FTIR and 2DIR spectra of the entire KcsA complex, which is the largest complex for which such modeling has been performed. We found that by combing isotope labeling with 2D IR spectroscopy, the signatures of potassium interaction with individual binding sites would be experimentally observable and identified specific labeling combinations that would maximize our expected experimental signatures. 
\end{abstract}
\pacs{XXXX, XXXX}
\submitto{\NJP}
\maketitle
\normalsize
\section{Introduction and background}
Transport processes are ubiquitous in biological systems. Their energetic, spatial and temporal scale ranges from that of entire cell populations down to the molecular and atomic scale. At each scale they are characterized by their high level of specificity, directedness and efficiency. Understanding the underlying principles of these processes and their relevance for biological function represents some the current frontiers of molecular biology, biochemistry and more recently that of physics. 
Due to the energy scale regime and high level of coupling to the environment, quantum coherence effects have been considered irrelevant in biological transport process, and such processes have been successfully described by rate equation models and classical thermodynamics. More recently, as best demonstrated by experiments on photosynthetic reaction centers, \cite{Engel07} there is evidence that for certain biological processes, the involved temporal, spatial, and energetic scales, permit non-classical phenomena such as coherent excitation transfer across protein complexes, and moreover might have provided an evolutionary advantage. There is evidence that these evolutionary advantages might be best harnessed for systems that have not evolved to be purely coherent nor classical, but rather at the cross section between the two regimes where an interplay between environmentally induced decoherence and quantum dynamics can be exploited \cite{Aspuru08,PlenioH08}. Although these effects have been rigorously studied theoretically \cite{CarusoCDHP09,ChinDCHP09, RebentrostMA09, Whaley} and experimentally for the photosynthetic system, it is a valid question to what extent some of these results can be extended to other biological systems. Specifically, if these effects are important, are they limited to the dynamics of electrons only, or can quantum coherence or tunneling, under certain circumstances where thermal fluctuations lead to sufficiently small potential barriers, also play a role in the transport of more massive particles such as ions? This is a controversial topic, on which even the authors share different opinions.
In this context, ion channel and pumps, due to the size, time and energy scales of the transport processes that they are involved in, \cite{Hille2001, Gadsby2009} represent another class of nano-scale ``protein machines'' where quantum coherence effects might play a functional role. Given our incomplete understanding of the mechanism that leads to their high level of specificity, directedness and efficiency, the observation of any quantum coherence in such systems may shed new light on the fundamental functional principles in such systems that might eventually be extended more broadly to other areas of biochemistry and molecular biology.
\vspace{0.25cm}
\subsection{Ion Channels and Anomalous Kinetics of Ion Translocation}
\begin{figure}[t]
\begin{center}
\centering
\includegraphics[width=8.3cm]{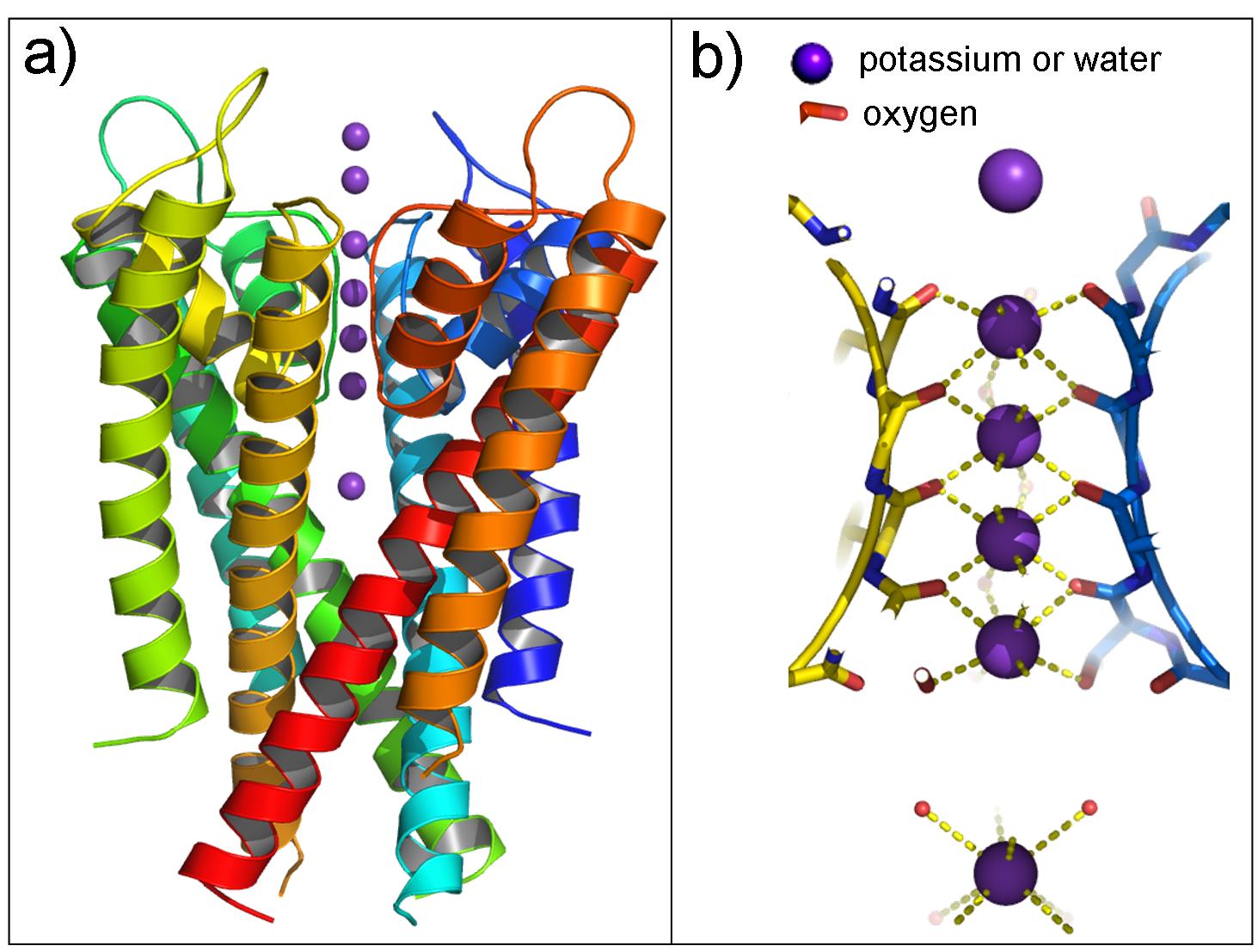}
\caption{Schematic illustration of the KcsA potassium channel after PDB 1K4C. KcsA protein complex with four transmembrane subunits (left) and selectivity with four axial trapping sites formed by the carbonyl oxygen atoms in which a potassium ion or a water molecule can be trapped.}
\label{kcsa_and_selectivity}
\end{center}
\end{figure}

One of the best studied classes of ion channels are $\mbox{K}^+$ channels \cite{Doyle98, Sokolova2001, Berneche2001, Noskov2004}. They play key roles in the shaping of action potentials used for neuronal signaling and cardiac muscle activity, ionic homeostasis, cell proliferation \cite{Pardo2004} and epithelial fluid transport \cite{O'Grady2003}. The structure of the $\mbox{K}^+$channel from the bacterium \textit{Streptomyces lividans} (KcsA) shows that the potassium channel assembles as a homotetramer of membrane-spanning protein subunits \cite{Doyle98}. While there are different gating mechanisms in different $\mbox{K}^+$ channels, the selectivity filter is conserved across all potassium channels. It is formed by a tetramer of transmembrane helices near the extracellular surface of the membrane. It is comprised of a short peptide loop (TTVGY) from each of the four helices\cite{Morais-Cabral2001}. The oxygen atoms of the carbonyl groups of each of these five amino acids are pointing towards the center of the pore forming five axial potential minima for trapping of positive charges (Fig.~\ref{kcsa_and_selectivity}). The initial crystallographic studies have revealed the presence of either a potassium ion or a water molecule at all times at each of these binding site. These studies also showed that the axial separation between each of these sites is ~0.24 nm and the width of the filter is ~0.3 nm \cite{Morais-Cabral2001}. This observation implied that the process of ion transport across the selectivity filter must occur in a single-file fashion. Moreover, the above dimensions of the selectivity filter require that each $\mbox{K}^+$ ion must shed its hydration shell before entering the channel. These realizations were particularly remarkable as the potassium channel combines very high throughput rates ($10^8$ ions /sec) \cite{Gouaux2005}, close to the diffusion limit, with a high ($1:10^{4}$) discrimination rate \cite{Doyle98} between potassium and sodium. It has been a challenge to fully explain the underlying molecular and atomic interactions that lead to this observation. This task has been particularly hampered by the lack of experimental studies that yield insights into atomic details of the transport process and the associated protein dynamics at time resolutions that are capable of following individual ions through the channel.
\vspace{0.25cm}
Nevertheless, based on crystallographic data and computational modeling, several models for the selectivity and the high transport rate have been developed. In this context the currently well accepted model for ion selectivity is based on the so called ``snug-fit'' model \cite{Bezanill1972, Roux2005, Noskov2006}. It suggests that the network of the residues are forming a spatial arrangement such that the dehydrated $\mbox{K}^+$ ions fit snugly into the filter with water-like coordination by the backbone carbonyl oxygen atoms, while the selectivity filter cannot distort sufficiently to coordinate the $\mbox{Na}^+$ ions \cite{Doyle98, Zhou2001}. Considering that the atomic radius of $\mbox{K}^+$ and $\mbox{Na}^+$ only differs by 0.38 $\textit{\AA}$, the validity of the snug fit model to explain the ion selectivity of the selectivity filter depends on a sustained, rigid geometry with sub-$\textit{\AA}$ngstrom precision in order to discriminate the two cations. However proteins, similar to most bio-molecular complexes are flexible structures subjected to significant temperature fluctuations with amplitudes on the order of 0.75 to 1.0 $\textit{\AA}$ \cite{Roux2005, Noskov2006, Gwan2007}. 
Similarly, the high throughput rate of the ions has been explained through a model commonly referred to as the ``knock-on'' mechanism which assumes that the ion-ion channel attraction outside of the selectivity filter and the and ion-ion repulsion inside the filter have compensating effects \cite{Morais-Cabral2001, Gouaux2005, Noskov2004, Berneche2001}. Basically, it is assumed that ions are electrostatically attracted to the channel by negatively charged residues. The approach of such an ion from one side of the doubly occupied selectivity filter leads to the subsequent exit of another ion on the other side of the channel. Although this mechanism is in principle possible, it was shown \cite{Roux2005} to require a highly delicate energetic balance between the ion-ion channel attraction and ion-ion repulsion for this mechanism to be able to explain the observed high throughput rates. However, numeric calculations show that the Columbic repulsion during the transport cycle can vary by several tens of kcal per molecule. Essentially the interplay between the thermally induced time-varying electrostatic potential of the carbonyl groups trapping the $\mbox{K}^+$ ions and the electrostatic back action of the ions leads to a temporally and spatially highly dynamic potential landscape.
\vspace{0.25cm}
This still leaves us with some of the same fundamental questions: How does a flexible structure like the selectivity filter achieve ion selectivity and high throughput at the same time? It is also known that two different possible sequences of $\mbox{K}^+$ and water in the selectivity filter, commonly referred to as the 1,3 and 2,4 states, exist for which two $\mbox{K}^+$ ions are within the selectivity filter at all times \cite{Morais-Cabral2001}. But what is the functional significance of these configurations? There is some evidence \cite{Roux2005, Noskov2006} suggesting that the thermal fluctuations could in fact be required for the channel function, so what are the underlying mechanisms for using such fluctuations for the channel function? It is now believed \cite{Roux2005, Noskov2006} that to understand ion selectivity and transport, one has to account for competing microscopic interactions for whose quantitative description, accurate models on the atomic level are required. 
\vspace{0.25cm}
Recently it was demonstrated theoretically that, based on the time and energetic scales involved in the selectivity filter, the ion selectivity and transport cannot be entirely a classical process but involves quantum coherence \cite{Vaziri2010a}. It was shown that the Hamiltonian describing the dynamics of the ``chain'' of water-ion molecules that is subjected to the axial Coulomb potential in the selectivity filter can be written as a coupled series of quantum harmonic oscillators connected to a source and a sink where excitations enter and leave the system. Using such a model and the parameters of the $\mbox{K}^+$ selectivity filter \cite{Gwan2007, Berneche2001}, it was demonstrated that the dynamics of the coupled system will exhibit constructive and destructive interferences if the system is driven by an external oscillating field with frequencies on the scale of the ion transport rate. These interferences lead to resonances in the transport efficiency at which the ion conduction is highly impaired, which sharply contrast the essentially frequency-independent results expected of a classical ``hopping type'' transport process. Although the suggested experiments based on quantum resonances could provide evidence for coherence in the dynamics of the underlying mechanism of the ion selectivity and transport it is certainly more informative to be able to study the ion transport process and the associated protein dynamics more directly at picosecond time scale.
\subsection{Ultrafast two-dimensional infrared vibrational spectroscopy}
Two-dimensional infrared spectroscopy (2D IR) is an emerging molecular spectroscopy for the biomedical and biophysical sciences. 2D IR characterizes molecular structure, provides information on components in heterogeneous samples, and has the high time-resolution required to measure fast kinetics \cite{Hamm1998,Tokmakoff2007,Hochstrasser2007, Zhuang2009}. Analogous to 2D NMR methods, 2D IR uses sequences of infrared pulses to excite and detect molecular vibrations. The spectra in this manuscript were acquired with short infrared pulses, thus the experiment was performed in the time-domain, and the resulting signal Fourier transformed. The resulting frequency-frequency 2D spectrum provides the ability to correlate different spectral resonances. Cross peaks can be used to enhance the information content of congested spectra, reveal the coupling or connectivity of the parts, or watch the time-dependent exchange of different molecular species. Since IR and Raman spectroscopy provide markers of molecular structure, characterizing cross-peaks can provide quantitative assays and in some cases, can be used to obtain precise structures or proximities. Since it makes these measurements with picosecond time resolution, it is a method of characterizing transient molecular structures and time-evolving molecular structure. Moreover, as cross-peaks arise from energy transfer pathways between two vibrational modes, beating in the amplitude of these cross peaks for different inter-pulse delay times provides evidence for the preparation of a coherent state.\cite{Deflores2006,Khalil2004} This phenomenon was used in the electronic 2D spectroscopy of the photosynthetic complex to show coherent excitation transfer. These beats, however, can originate for either quantum mechanical and classical couplings.\cite{Zimanyi2010}
\subsection{A Brief Guide to Reading 2D IR Spectra}
\subsubsection{Peak Positions}
A 2D IR spectrum is typically plotted as a two-dimensional contour surface as a function of excitation ($\omega_1$) and detection ($\omega_3$) frequencies. The excitation frequencies of all of the peaks, or their $\omega_1$ position, can be compared to the linear absorption spectrum since this describes the possible excitation frequencies of the system from the ground state to the first excited state ($\left|0\right> \rightarrow \left|1\right>$ quantum transitions). This can be observed by comparing the FT IR spectrum and the $\omega_1$ 2D IR peak positions in Figure~\ref{spectroscopy_model_compounds}. A diagonal peak ($\omega_1=\omega_3$) arises from coherence that remains in the same mode for the excitation and detection periods, which can be assigned to chemically distinct vibrational normal modes or one-quantum eigenstates. An off-diagonal peak at $(\omega_1^*,\omega_3^*)$ indicates that induced or spontaneous transfer of energy from the coherence prepared at $\omega_1^*$ during the excitation period to $\omega_3^*$ during the detection period. This can arise from vibrational coupling between these two modes.\cite{Khalil2003} In Figure ~\ref{spectroscopy_model_compounds}, only diagonal peaks are observed, indicating weak or negligible coupling on the timescale of this experiment. Generally, weak coupling can be probed by allowing the system to evolve by increasing the delay between excitation and detection periods to give time for slower processes to distribute energy amongst the observable modes.
\subsubsection{Positive and Negative Peaks}
In all of the 2D IR spectra in this manuscript, such as Figure~\ref{spectroscopy_model_compounds}, red and blue shaded contours indicate positive and negative features. Since the emitted 2D IR signal is recorded interferometrically by combining it with a local oscillator, signal electric field amplitudes that are in-phase or $\pi$ out-of-phase interfere with the local oscillator give rise to positive- or negative-signed peaks after Fourier transformation. These phases arise result from interferences due to $\left|1\right> \rightarrow \left|0\right>$ quantum transitions or $\left|1\right> \rightarrow \left|2\right>$ quantum transitions, respectively chosen to be positive and negative phase. This means that for each isolated oscillator, a doublet of positive and negative peaks will arise, shifted in the $\omega_3$ direction by the vibrational anharmonicity, or energy difference between the 2-1 and 1-0 energy gaps. For a harmonic system, these two transitions perfectly cancel and no 2D IR signal is observed.
\subsubsection{Lineshapes}
For an ensemble molecular oscillators that undergo no frequency fluctuations and whose relaxation takes the form of a single exponential, the 2D IR signature will be a two-dimensional Lorentzian lineshape. (One example is the peak in Figure~\ref{spectroscopy_model_compounds}h at $\omega_1=\omega_3=1750\ cm^{-1}$.) Dephasing processes such as frequency fluctuations due to interactions with the bath and population relaxation that are fast compared to the time it takes for the molecular oscillator to define its frequency ($\approx10T$, where $T$ is the oscillation period, which is 20 fs for an oscillator at $1666 cm^{-1}$), will broaden the line but maintain its Lorentzian shape. Strong coupling or coherence transfer events will also affect 2D IR lineshapes, necessitating detailed models to interpret the lineshapes rigorously. Slower fluctuations will induce correlations between the $\omega_1$ and $\omega_3$ frequencies, resulting in a diagonally elongated lineshape, which is often fit to a Gaussian lineshape. (See the peak at $\omega_1=\omega_3=1675\ cm^{-1}$ in Figure~\ref{spectroscopy_model_compounds}g) Frequency fluctuations that are much slower than the timescale of the experiment (the observable window is set by the vibrational lifetime, which destroys vibrational coherences, and is typically 1-10 ps) can be viewed as static disorder. For fluctuations occurring within this time window, modeling 2D IR lineshapes can reveal information about the system-bath coupling or interchange between molecular configurations causing the frequency changes. To many biologically relevant processes, this is considered a short-timescale window and here 2D IR can be used to provide a snapshot for evolving structure. For the thesis that quantum coherence plays a role in the ion transfer properties of KcsA, sensitivity to vibrational coherence on femtosecond timescales is a key strength of 2D IR spectroscopy.
\vspace{0.25cm}
The potential of 2D IR spectroscopy has been most clearly seen with studies of proteins and peptides using amide I spectroscopy. The amide I vibration ($1600-1700\ cm ^{-1}$, primarily CO stretch) is sensitive to the type and amount of secondary structures and not strongly influenced by side chains \cite{Jackson1995, Byler1986}. $\beta $-sheets have a strong absorption band at $1630-1640\ cm^{-1}$ that shifts with sheet size and a weaker band at $~1680\ cm^{-1}$. The $\alpha$-helix and random coil structure are located at $1650-1660\ cm^{-1}$ and $1640-1650\ cm^{-1}$, respectively. This sensitivity results because amide I vibrations of proteins are delocalized over secondary structures of the protein, as a result of strong couplings between amide I oscillators. The spectroscopy of these ``excitonic'' states reflects the size of secondary structure and the underlying structural arrangement of oscillators, but lack local (site-specific) detail.
To extract site-specific information without perturbing the protein, isotope labels can be introduced into the amide I carbonyl \cite{Torres2001, Decatur2006, Manor2009}. Labeling the peptide carbonyl with $^{13}\mbox{C}$ and / or $^{18}\mbox{O}$ provides a red-shift between 35 and 65 $cm^{-1}$ from the main amide I band, thereby spectrally isolating a particular peptide unit. This strategy has proven very useful in 2D IR for interrogating site specific peptide and protein structure. Because an amide I unit is sensitive to hydrogen bonding ($\approx 20\ cm^{-1}$ / H-bond to C=O), an isotope label provides an excellent measure of the local solvent exposure. Additionally, site-specific contacts can be monitored through the introduction of a pair of isotope labels. The frequency of the two oscillator coupled state provides distance sensitivity over a range of $\approx 1-4\ \AA$, allowing for precise monitoring of contacts. 
Cross-peaks in 2D spectra contain the relative orientation and coupling between vibrational dipoles which can be translated to a structure with a molecular model \cite{Khalil2003}. 
Amide I spectroscopy has proven particularly attractive because structure-based spectroscopic models exist that allow amide I IR spectra to be modeled from a protein structure \cite{Ganim2006,Jansen:2009p896, Marai2010}. For a proposed structure or pathway with atomistic structures, 2D IR data can be calculated and compared to the experiment to confirm or reject the proposed structure. A hierarchy of approximations exist for calculating 2D IR spectra. The most accurate models in use simulate the time-evolution of the system of molecular oscillators and extract trajectories for all relevant frequencies and transition dipoles. The quantum mechanical propagators describing the density matrix evolution during a 2D IR experiment are reconstructed and the signals resulting from all pathways induced by the excitation and detection pulses are summed, and ensemble-averaged. A substantial improvement is gained by assuming that slow frequency evolution is much slower than the timescale of the experiment. This so-called "static averaging approximation" has been used to calculate the 2D IR spectra of large protein systems suitable to predict and interpret experimental data. Finally, \textit{ad hoc} (rather than structure-based) models exist for describing 2D IR spectra in terms of distributions of frequencies and molecular couplings.\cite{Deflores2009, Wang2004, Maekawa2008, Krummel2006}
All these properties make 2D IR an emerging tool for studying dynamic interactions in protein complexes such as transmembrane proteins \cite{Mukherjee2006,Fang2006}. This is particularly the case when isotope labels such as $\mbox{C}^{13}$ and / or $^{18}\mbox{O}$ are combined with 2D IR to achieve high temporal and structural resolution.\cite{Andersen2009} As a first step towards elucidating the function of the selectivity filter we have experimentally identified different spectroscopic signatures of $\mbox{K}^+$ binding in model compounds that represent the signatures of $\mbox{K}^+$ binding for different conformational states of the selectivity filter. Further, based on our model for amide I spectroscopy we have shown the expected shifts in the $\mbox{K}^+$ binding signatures in the selectivity filter when isotope $\mbox{C}^{13}$ and / or $^{18}\mbox{O}$ labeled carbonyl are used. 
\section{Vibrational spectroscopy of $\mbox{K}^+$ binding model compounds: vibrational excitons as probes of $\mbox{K}^+$ binding in the selectivity filter}
\begin{figure}[t]
\centering
\includegraphics[width=13cm]{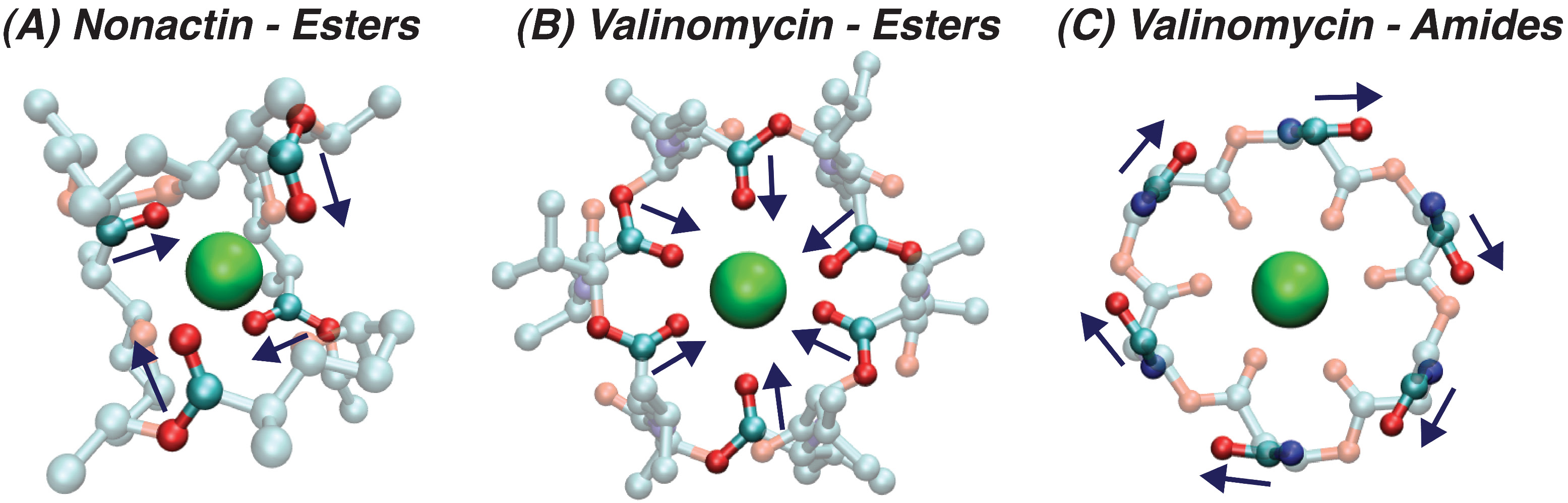}
\begin{center}
\caption{Molecular structure of $\mbox{K}^+$ binding model compounds nonactin and valinomycin coordinating potassium, with ester and amide groups highlighted and according transition dipole directions indicated with arrows. (A) Potassium is eight-fold coordinated in nonactin by four ester carbonyl oxygens and four ester oxygens. (B) Potassium is octahedrally (sixfold) coordinated by the valinomycin ester carbonyl oxygens. (C) The valinomycin amide groups point axially and do not coordinate potassium (side chains removed for clarity). Structures obtained from Refs.~\cite{Kilbourn1967,Neupert-Laves1975} Structured rendered with VMD\cite{Humphrey1996} and POV-Ray\cite{povray}.}
\label{model_compounds}
\end{center}
\end{figure}

The binding of $\mbox{K}^+$ ions by peptide units within the selectivity filter requires conformational changes leading to eight or six-fold coordination of the $\mbox{K}^+$ ion. Figure~\ref{selectivity_filter} shows five planes of oxygen atoms responsible for coordination. Eight-fold coordination comes about when a $\mbox{K}^+$ sits between two planes, where it is electrostatically coordinated between four carbonyls from each plane. For six-fold coordination, the $\mbox{K}^+$ sits within a ring of four carbonyl oxygens, and is further coordinated from above and below by two water molecules. Similar coordination symmetries, although with different chemical motifs, are also observed in certain antibiotics responsible for diffusive transport of ions across bacterial cell membranes. We have used valinomycin and nonactin in this respect as models for biological $\mbox{K}^+$ binding (Fig.~\ref{model_compounds}). Nonactin, in particular, is the closest in structural similarity to a single $\mbox{K}^+$ binding site within the KcsA selectivity filter. These molecules therefore serve as valuable model compounds to study FT IR, Raman \cite{Asher1974,Asher1977} signatures of $\mbox{K}^+$ binding, and this will be extended to 2D IR. The strong electrostatic interaction of the ion with carbonyls, the close proximity and coupling of those carbonyls, and the high symmetry normal modes that arise from binding are all indicators that vibrational spectroscopy will see strong excitonic resonance enhancements and pronounced variation in IR and Raman spectra due to selection rules. To validate the suitability of this technique for studying the $\mbox{K}^+$ binding in the selectivity filter in real time, we have performed FT IR, Raman and 2D IR spectroscopy on valinomycin and nonactin.
\subsection{Infrared and Raman Spectroscopy Experimental Methods}
\small
The FT IR spectra were acquired using a home-built brass cell equipped with 1 mm thick CaF$_2$ windows and a 50 $\mu m$ thick Teflon spacer\cite{Smith:2008} with a Nicolet 380 FT IR spectrometer. In all experiments, the solvent was methanol-d4 (Cambridge Isotopes, Cambridge MA). Valinomycin and nonactin were purchased from A.G. Scientific Inc. (San Diego, CA) and used without further purification. The valinomycin samples were 13 mM with 0 mM KCl and 4.5 mM with 30 mM KCl. In the absence of KCl, nonactin is only slightly soluble in methanol. Thus, to obtain the nonactin spectra in 3 mM and 100 mM KCl, methanol with the desired salt concentration was added to an excess of nonactin to create a saturated solution. This solution was centrifuged and the spectra of the supernatant were measured.
The 2D IR spectra were acquired using IR pulses that were centered at ~1700 cm$^{-1}$, using methods that have been described elsewhere.\cite{Chung2009} The signals were heterodyne-detected by overlap with a local oscillator and acquired with a spectrometer providing 2.0 cm$^{-1}$ resolution in the detection ($\omega_3$) axis. The spectra were acquired as a set rephasing and non-rephasing experiments, as a function of coherence times ($\tau_1$) spanning $\approx$ 5 ps in 4 fs steps. The waiting time ($\tau_2$) was zero in all spectra. The spectra were phased by comparing the zero-time slice to the two-beam pump-probe spectra. Subsequently, the rephasing and non-rephasing spectra were Fourier transformed and summed. The polarizations of all excitation pulses and local oscillator were mutually parallel to provide all parallel 2D IR spectra.
\normalsize
\subsection{Infrared and Raman Spectra of Model Compounds and Ion Binding}
\begin{figure}[htbp!]
\centering
\begin{center}
\includegraphics[width=19.0cm,angle=90]{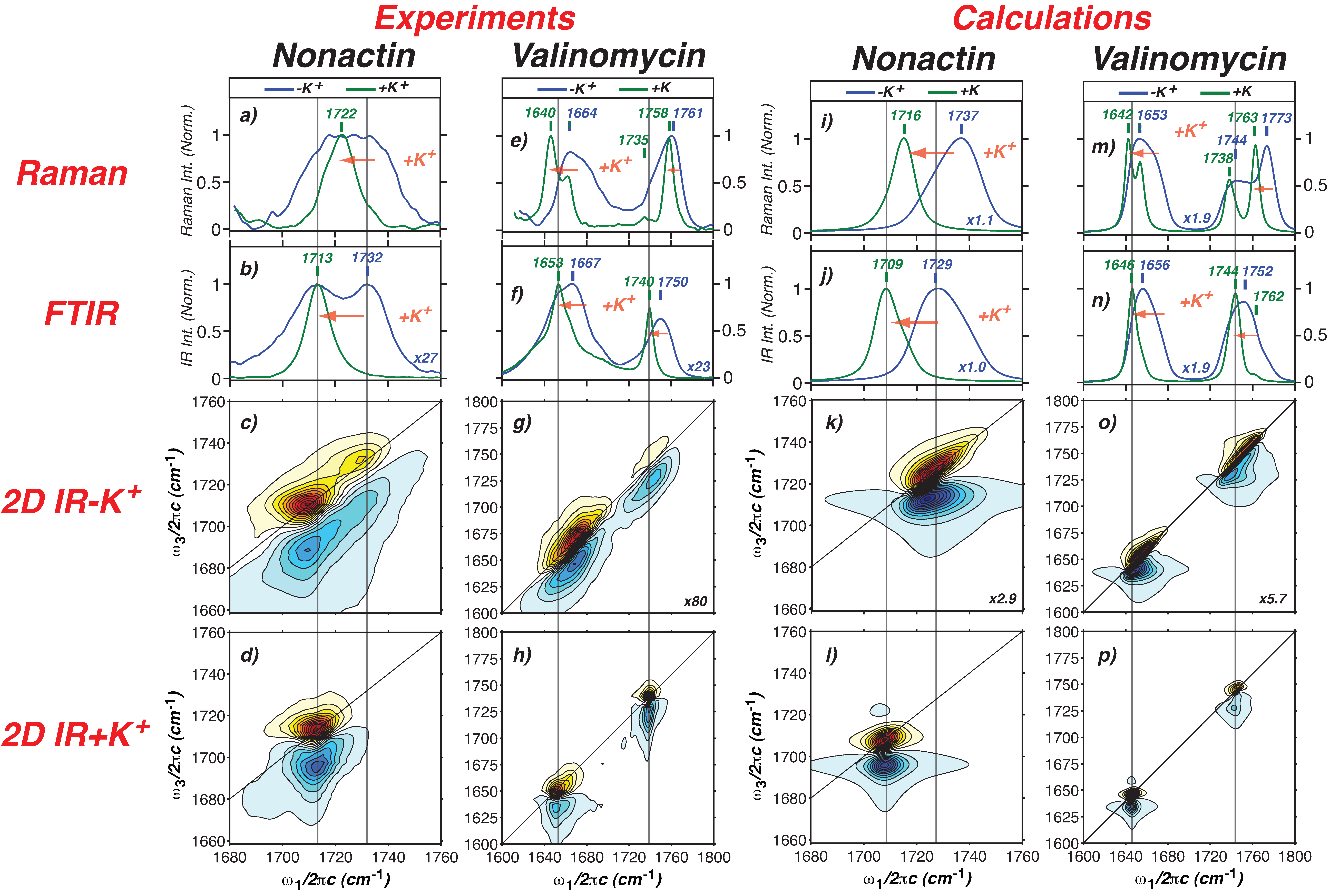}
\caption{Experimental (a-h) and calculated (i-p) FT IR, Raman, and 2D IR spectra of nonactin and valinomycin in the carbonyl stretching region with and without $\mbox{K}^+$. The Raman spectra in panels (a) and (e) were digitized from previously measurements.\cite{Asher1974,Asher1977}. In the 2D IR spectra, contours are drawn at 25 equally spaced levels. Peak positions are labeled and orange arrows indicate the trend with $\mbox{K}^+$ addition. All intensities were normalized; in each case, the signal increased with KCl addition when unambiguously measured under identical conditions, and thus the unbound spectra were scaled by the factors indicated. Overall, the spectra show narrowing and frequency red-shifting upon $\mbox{K}^+$-binding, due induced ordering and changes in the electrostatic environment.}
\label{spectroscopy_model_compounds}
\end{center}
\end{figure}

Figure~\ref{spectroscopy_model_compounds} shows the FT IR, Raman, and 2D IR spectra of valinomycin and nonactin with and without KCl. Nonactin and valinomycin both coordinate the $\mbox{K}^+$ with ester groups (stretching absorption at 1700-1800 $cm^{-1}$), and valinomycin additionally contains amide groups (1600-1700 $cm^{-1}$). The goal of considering all of these spectra simultaneously is to understand the structural changes induced by $\mbox{K}^+$ binding on the model compounds by comparing peak shifts, IR/Raman band splittings, and line shapes across all the observed carbonyl stretching bands. These vibrational signatures can, in turn, be used to design experiments and interpret data on the KcsA filter.
\newpage
\begin{figure}[htbp!]
\centering
\begin{center}
\includegraphics[width=13.0cm]{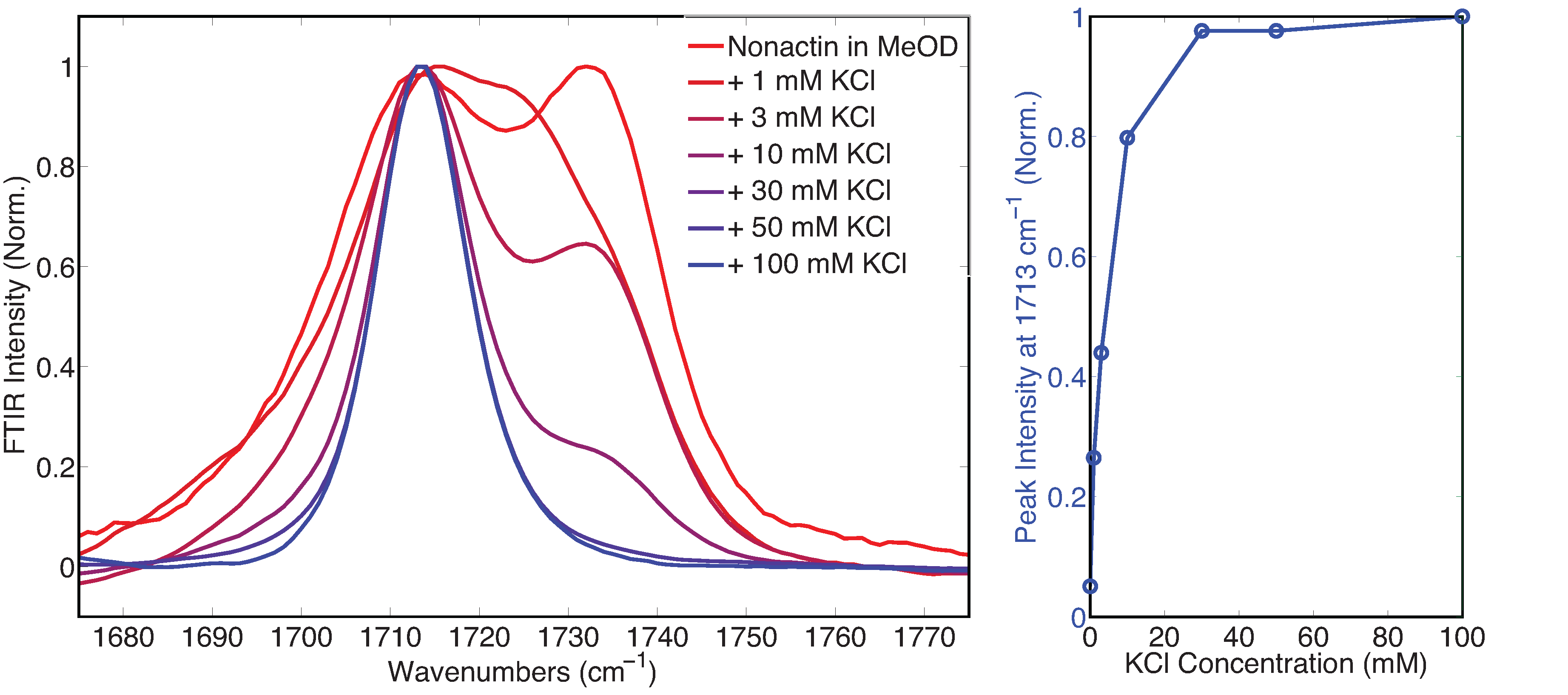}
\caption{KCl-concentration dependent nonactin FTIR spectra (a) show the growth of a peak at 1713 $cm^{-1}$ upon $\mbox{K}^{+}$ binding (normalized spectra). The intensity at 1713 $cm^{-1}$ as a function of concentration is shown in panel (b). Residually bound potassium is observed even with pure MeOD solvent.}
\label{conc_dep_non}
\end{center}
\end{figure}
\subsubsection{Peak Shifts Due to $\mbox{K}^+$ Binding}
A uniform feature across all of the spectra is a red-shift of the peaks upon $\mbox{K}^+$ binding. It is known that carbonyl stretching frequencies are sensitive to the local electrostatic environment. Since its oxygen atom carries a partial negative charge (i.e., -0.5$e$ according to the OPLS\cite{opls} molecular dynamics force-field), carbonyls preferentially coordinate positive charges, and in doing so, weaken the C=O bond and red-shifts the carbonyl stretching frequency. The FT IR and Raman spectra of Fig.~\ref{spectroscopy_model_compounds} are all consistent with this interpretation. Each of the valinomycin amide and ester peaks shifts $10-13\ cm^{-1}$ to the red. In the FT IR spectrum of nonactin, there is evidence for a two-peak structure in the unbound state, which collapses to one peak with $\mbox{K}^+$. The appearance of the bound nonactin peak in the KCl-free spectrum is due to incomplete removal of K$^{+}$ in the supplied sample. This is demonstrated in Figure~\ref{conc_dep_non}a, which shows that at 100 mM KCl, only the bound species is present (1713 $cm^{-1}$); at lower concentrations, an unbound peak appears at 1732 $cm^{-1}$, but the bound species does not fully disappear, even when no KCl is added (Fig.~\ref{conc_dep_non}b). The Raman spectrum of nonactin shows a narrowing upon adding $\mbox{K}^+$, which can be interpreted as a transition from a disordered ensemble to an ordered one. Since there is more of an intensity loss on the blue side, this can be interpreted in analogy to the FT IR as a decrease in the amount of unbound species, although there is no clear evidence for two peaks.
All of these peak shifts upon $\mbox{K}^+$ binding are more clearly reflected in the 2D IR spectra. For nonactin, the two peak structure seen in the FT IR is much better resolved with 2D IR. (Moreover the lack of a cross-peak between the two nonactin peaks is further evidence that these arise from separate species, e.g., a $\mbox{K}^+$-bound and unbound state, and not different vibrational modes of the same molecule.) The increased resolution is due to fact that 2D IR peaks scale as $\left | \mu\right|^4$ compared to $\left | \mu\right|^2$ in FT IR, which emphasizes those peaks arising from strong transition dipoles relative to peaks arising from a sum of weak transition dipoles. For example, this reduces the baseline between the two nonactin peaks. (This effect may also cause 2D IR peaks to be slightly shifted relative to FT IR absorptions.)
\subsubsection{Comparison of FT IR and Raman Bands}
For complexes with an inversion center, such as valinomycin, a vibrational mode will be either infrared or Raman active\cite{Harris1989} due to the respective odd or even transformations of the dipole vector (x, y, z) or polarizability tensors (x$^2$, xy, x$^2$-y$^2$, etc) in these point groups. Conceptually, the coupling between carbonyl groups results in eigenstates that are either symmetric under inversion (thus, the sum of the transition dipole vectors of constituent C=O groups is zero, with non-zero polarizability) or anti-symmetric under inversion (non-zero overall transition dipole, but canceling polarizabilities). Therefore, changes in the peak positions between IR and Raman spectra provide evidence for inversion symmetry, and the splitting between IR and Raman bands reports on the magnitude and sign of coupling between carbonyl groups.\cite{Krishnan1968,Sawyer1958}
In the absence of $\mbox{K}^+$, the IR and Raman spectra generally resemble one another, for both valinomycin and nonactin. This is consistent with a disordered unbound state that does not preserve the symmetry shown in Fig.~\ref{model_compounds}. Upon $\mbox{K}^+$ addition, it can clearly be seen that all of the IR and Raman peaks are in different positions, which indicates an ordered, symmetric structure. There is no \textit{a priori} pattern to the shifting between IR and Raman peaks: In nonactin the Raman spectrum peaks at a higher frequency than the FT IR. In the amide region of valinomycin, the FT IR band peaks in between a two-peak structure in the Raman spectrum. In the ester region of valinomycin, the FT IR peaks at a lower frequency than the Raman band. Detailed modeling that incorporates the relative orientations of C=O groups and the coupling between them must be used to interpret these changes.
\subsubsection{Line Shape Changes}
In addition to the peak shifts, all of the FT IR and Raman bands narrow upon $\mbox{K}^+$ addition. Observing these changes in the 2D IR spectra gives strong support to the conjecture that adding $\mbox{K}^+$ causes a shift from a structurally disordered ensemble to an ordered one. In valinomycin, the diagonal narrowing is clear in both the amide and ester regions. For nonactin, the unbound spectrum again appears to be a two-component mixture, one peak of which is nearly identical in position and line shape to the $\mbox{K}^+$-bound state.
For valinomycin, no cross-peaks are clearly visible, indicating weak coupling between the amide and ester modes. This implies that these two regions of the spectrum are relatively uncoupled on the timescale of the experiment (compared to the energy difference of 100 cm$^{-1}$). (Cross-peaks are found to grow in with waiting time, and become clear at ~4 ps; unpublished data.)
\subsection{Modeling Infrared and Raman Spectra of Model Compounds}
Most vibrational spectral changes upon ion complexation can be described in terms of a simple spectroscopic model based on the symmetry of the ion-complexed model compounds and the addition of structural disorder. Such a model can be used to quantify the peak shift due to $\mbox{K}^+$ binding when it appears concomitantly with structural changes to the ensemble, to quantify the increase in disorder when $\mbox{K}^+$ is released, and to estimate the amount of coupling between the carbonyl modes. Using the symmetry properties of the published structures of $\mbox{K}^+$-bound valinomycin and nonactin can be used to reduce the number of variable parameters in the model.
It is assumed that the Hamiltonian has an excitonic structure; namely that the amide and ester carbonyl stretching modes form vibrationally uncoupled manifolds (a suitable approximation given the lack of cross-peaks between these modes), and that eigenstates of each manifold are formed from linear combinations of local amide or ester carbonyl stretching modes. Such models have been routinely used to model vibrational spectra of peptides and proteins,\cite{Hamm1998,Measey2006,Zhuang2009,Ganim2006,Wang2004} and are conceptually equivalent to tight-binding models in solid-state physics.\cite{Slater1954} To calculate vibrational spectra from this excitonic model, one needs a set of transition dipoles and transition polarizabilities for the local amide and ester modes to be modeled. One also needs a Hamiltonian that describes the energy of each local amide or ester site and the coupling among them.
\subsubsection{Transition Dipoles and Polarizabilities}
The crystal structure of valinomycin-$\mbox{K}^+$ shows that the ester and amide groups individually belong to the $S_6$ point group.\cite{Harris1989} As shown in Fig.~\ref{model_compounds}, the six ester carbonyl oxygens point radially inward and the six amide carbonyl oxygens point equatorial. The four ester groups of nonactin belong to the $D_{2d}$ point group. From these structures, the (x,y,z)-components of the bond vectors for the carbonyls can be described as,
\begin{eqnarray}
\textrm{valinomycin amide: } \hat{r}^{CO}_n &=& \left( \begin{array}{c} cos(n\pi/3) \\ sin(n\pi/3)\\ 0 \end{array} \right), n=0..5\\
\textrm{valinomycin ester: } \hat{r}^{CO}_n &=& \left( \begin{array}{c} sin(n\pi/3) \\ cos(n\pi/3)\\ 0 \end{array} \right), n=0..5, \textrm{and} \\
\textrm{nonactin ester: } \hat{r}^{CO}_n &=& \left( \begin{array}{c} cos(n\pi/2) \\ sin(n\pi/2) \\ 0 \end{array} \right), n=0..3,
\end{eqnarray}
where the $z$ axis is taken to be in the direction relating rotational symmetry, and $x$ and $y$ are defined for convenience based on the first carbonyl. (Since only the magnitudes of the ensemble-averaged eigenstate transition dipoles are observed, the origin is inconsequential.)
The transition dipole operator, relevant for calculating infrared spectra, for each amide or ester unit is assumed to lie along the C=O bond vector, 
\begin{equation}
\vec{\mu}_n \propto \hat{r}^{CO}_n.
\end{equation}
The transition polarizability operator, relevant for calculating Raman spectra, for each amide or ester unit describes on-axis polarizability changes,
\begin{equation}
\vec{\rho}_n \propto \left( \begin{array}{ccc} (r^{CO}_{x,n})^2 & 0 & 0 \\ 0 & (r^{CO}_{y,n})^2 & 0 \\ 0 & 0 & (r^{CO}_{z,n})^2 \end{array} \right).
\end{equation}
Because only relative intensities are being modeled, the proportionality constants are not needed.
\subsubsection{Hamiltonian}
The symmetry of valinomycin indicates, in the absence of any disorder, that the Hamiltonian should have the structure,
\begin{equation}
H^0 = \left( \begin{array}{cccccc} \epsilon & \beta_{12} & \beta_{13} & \beta_{14} & \beta_{13} & \beta_{12} \\
\beta_{12} & \epsilon & \beta_{12} & \beta_{13} & \beta_{14} & \beta_{13} \\
\beta_{13} & \beta_{12} & \epsilon & \beta_{12} & \beta_{13} & \beta_{14} \\
\beta_{14} & \beta_{13} & \beta_{12} & \epsilon & \beta_{12} & \beta_{13} \\
\beta_{13} & \beta_{14} & \beta_{13} & \beta_{12} & \epsilon & \beta_{12} \\
\beta_{12} & \beta_{13} & \beta_{14} & \beta_{13} & \beta_{12} & \epsilon \end{array} \right),
\end{equation}
where $\beta_{12}$ is the nearest neighbor coupling, $\beta_{13}$ is the next-nearest neighbor coupling, etc. The parameter $\epsilon$ determines the frequency for the uncoupled carbonyl stretching modes, and varies based on the covalent bonded structure of each molecule, but each carbonyl in the molecule is chemically equivalent and has the same $\epsilon$. To simulate the electrostatic shift of $\mbox{K}^+$-binding, an additional parameter, $\Delta\epsilon_{bound}$, was added such that $\epsilon_{bound}=\epsilon_0+\Delta\epsilon_{bound}$. In general, each of these would be a fit parameter in the model, but observation of the molecule structure can be used to reduce this Hamiltonian to two parameters.
For the ester groups, an estimation of the vibrational couplings using the transition dipole coupling (TDC) model\cite{Moore1975} yields similar values for each of the couplings. The TDC model models the coupling between sites as dipole-dipole coupling, and thus takes as input parameters the magnitude of the dipoles, separation, and orientation. Using only the distance and orientation factor to scale the relative couplings gives: $\beta_{12} : \beta_{13} : \beta_{14} \approx 1.0 : 0.9 : 0.4$. For the amide groups, the distances between 1-3 and 1-4 neighbors are large (8.2 and 9.7 $\AA$, respectively), and a TDC calculation supports the assumption that $\beta_{12}$ is the only nonzero coupling ($\beta_{13}=\beta_{14}=0$). ($\beta_{12}$ is sufficiently small that constraining the amide couplings as for the esters makes no significant difference in the spectra.) Thus for each vibrational manifold, $\beta_{12}$ (hereafter called $\beta$) is the only fitting parameter to describe vibrational coupling.
For nonactin, the symmetry dictates that, in the absence of disorder, the Hamiltonian should have the structure,
\begin{equation}
H^0 = \left( \begin{array}{cccccc} \epsilon & \beta_{12} & \beta_{13} & \beta_{12}\\
\beta_{12} & \epsilon & \beta_{12} & \beta_{13}\\
\beta_{13} & \beta_{12} & \epsilon & \beta_{12}\\
\beta_{12} & \beta_{13} & \beta_{12} & \epsilon \end{array} \right).
\end{equation}
Each ester has two bonded, nearest-neighbors and one opposing group. Note that inversion symmetry is not present in nonactin like it is in valinomycin, indicating that the vibrational manifold may not be separated into IR-active and Raman-active components. Similar distance and orientation considerations as in valinomycin allows one to assume that $\beta_{12}=0$. Thus $\beta_{13}=0$ (hereafter called $\beta$) is the only coupling fit parameter.
Note that these symmetry relations are derived from the structure of $\mbox{K}^+$-bound valinomycin and nonactin, and may not hold for the unbound states. In this modeling, it is assumed for simplicity that the same average couplings remain in the unbound state, but that a large amount of disorder is added.
\subsubsection{Disorder}
The 2D IR spectra allow for estimation of the amount of structural disorder by observing the diagonal elongation in each peak. Increasing structural disorder in the vibrational chromophores leads to lack of a consistent angle or distance between carbonyl groups, which changes the coupling value. In a flexible, potassium-free state, the carbonyl groups interact more with the solvent, leading to larger variations hydrogen bonding configurations, and therefore, in the local mode frequencies of the carbonyl groups. This type of disorder, which was important to include in both the unbound and ion-bound state, was modeled by summing an ensemble of spectra generated from Hamiltonians allowing for variations in the parameters from an uncorrelated, Gaussian distribution,
\begin{eqnarray}
p(H_{ij}) = \frac{1}{\sigma_H \sqrt{2 \pi}} exp\left[-\frac{(H_{ij}-H^0_{ij})^2}{2 \sigma_H^2}\right]
\end{eqnarray}
The spectra resulting from each disorder realization (averaged until convergence, or 5000 realizations) was simulated as a sum over eigenstates ($N=6$ for valinomycin and $4$ for nonactin),
\begin{eqnarray}
S^{IR}(\omega) &=& \sum_n^{N} \frac{-\left|\mu_n\right|^2}{\omega-\epsilon_n + i\gamma} \\
S^{Iso\ Ram}(\omega) &=& \sum_n^{N} \frac{-1}{9}\frac{Tr\left[\rho_n\right]^2}{\omega-\epsilon_n + i\gamma} \\
S^{Aniso\ Ram}(\omega) &=& \sum_n^{N} \frac{-1}{\omega-\epsilon_n + i\gamma} \nonumber \\
&& \bigg[ \frac{1}{2}\left(\left(\rho_n^{xx}-\rho_n^{yy}\right)^2+\left(\rho_n^{yy}-\rho_n^{zz}\right)^2+\left(\rho_n^{zz}-\rho_n^{xx}\right)^2\right) \nonumber \\
&& +\frac{3}{4}\left(\left(\rho_n^{xy}+\rho_n^{yx}\right)^2+\left(\rho_n^{yz}+\rho_n^{zy}\right)^2+\left(\rho_n^{xz}+\rho_n^{zx}\right)^2\right)\bigg].
\end{eqnarray}
was summed to give the disorder averaged spectra plotted in Fig. \ref{spectroscopy_model_compounds}. All Raman spectra plotted are the unpolarized spectra obtained by
\begin{equation}
S^{Unpol.\ Ram} = S^{Iso.\ Ram}+\frac{10}{3}S^{Aniso.\ Ram}.
\end{equation}
The 2D IR spectra were obtained by scaling the one-quantum Hamiltonian using an anharmonicity of 16 cm$^{-1}$ and performing the sum over disorder realizations as previously described.\cite{Ganim2006}
\subsection{Fit Results}
\begin{table}[htdp]
\begin{center}
\begin{tabular}{|c|c|c|c|}
& Valinomycin Amide & Valinomycin Ester & Nonactin\\
\hline
$\epsilon_0$ ($cm^{-1}$) & 1661 & 1752 & 1732 \\
$\Delta\epsilon_{bound}$($cm^{-1}$) & -11 & -7 & -20 \\
$\beta$ ($cm^{-1}$) & -3.5 & 4.1 & 3.0 \\
$\sigma_{H,bound}$ ($cm^{-1}$) & 1.5 & 2 & 2 \\
$\sigma_{H,unbound}$ ($cm^{-1}$) & 5 & 6 & 4.5 \\
$\gamma$ ($cm^{-1}$) & 3 & 3 & 3 \\
\end{tabular}
\end{center}
\label{fitparams}
\caption{Fit parameters for the spectroscopic model to the Raman, FTIR, and 2DIR spectra of valinomycin and nonactin with and without $\mbox{K}^+$.}
\end{table}%
Each parameter in Table \ref{fitparams} has an associated observable in the IR and Raman spectra of bound and unbound states. The overall position of the bands were determined by $\epsilon_0$, and the shift upon $\mbox{K}^+$ binding was determined by $\Delta\epsilon_{bound}$. The splitting between IR and Raman peaks was determined by $\beta$. The parameter $\gamma$ was set by observing the anti-diagonal linewidth of the 2D IR peaks, which is relatively insensitive to broadening induced by disorder. The parameters $\sigma_{H,bound}$ and $\sigma_{H,unbound}$ affected the linewidths in the FT IR spectra and Raman spectra, as well as the diagonal linewidth of the 2D IR spectra, and was set to fit all of these observables.
The ensemble of structural configurations that give rise to disorder in the Hamiltonian necessarily also causes disorder in the orientations of transition dipoles and transition polarizabilities. When this was included in the model, its effect was to cause the IR and Raman spectra to more closely resemble one another, as expected due to the reduction of symmetry. However, since its inclusion made negligible difference to the fits, it was omitted for simplicity.
The calculated spectra (Fig. \ref{spectroscopy_model_compounds}) reproduce most of the trends seen in nonactin and valinomycin upon $\mbox{K}^+$ binding- peak shifts and broadening. For valinomycin, the pattern of splitting between the FT IR and Raman peaks is reproduced. In the amide and ester regions, the frequency order of IR and Raman peaks is reproduced. Also, as in the experimental spectra, the peaks narrow and redshift upon $\mbox{K}^+$ binding. Since the nonactin KCl-concentration-dependence (Fig.~\ref{conc_dep_non}) indicates the presence of a two-peak structure, these experimental spectra can be understood as combinations of the calculated bound and unbound spectra. While the splitting between the FT IR and Raman peak is reproduced in the ordered state, the experimental Raman spectrum shows an additional peak shift that was not reproduced in the calculation. This can arise from inconsistent experimental conditions between the FT IR and Raman spectra or if the Raman spectrum of bound nonactin changes at low $\mbox{K}^+$ concentrations.
In the 2D IR spectra of valinomycin, the unbound state shows diagonally elongated peaks, which redshift and narrow in the bound state. All of the calculated valinomycin and nonactin spectra show $\omega_3$-extension of the overtone peaks, as seen in the experimental spectrum. This appears as a result of interference with cross-peaks. For example, in valinomycin the amide doublet interferes with the negative lobe of a cross-peak to a resonance at $1680\ cm^{-1}$, which depletes from the fundamental and enhances the overtone, causing the apparent $\omega_3$-extension. The valinomycin ester undergoes such an interference with a weak peak at $1764\ cm^{-1}$. The same effect is also present in the nonactin spectrum, however to a much weaker extent.
\subsection{Conclusions for Model Compounds}
The fitted spectral parameters can be used to build an intuitive picture of the structural changes to the nonactin and valinomycin ensembles with and without added $\mbox{K}^+$ that is consistent with all of the observed spectra. In solution, the $\mbox{K}^+$-bound valinomycin resembles the crystal structure, although with a small amount of disorder. The amount of disorder can be judged by comparing the variation in Hamiltonian elements to the coupling ($\left|\beta\right| > \sigma_{H,bound}$ for the amide and ester groups from Table~\ref{fitparams}). In the large disorder limit, the fluctuations, $\left|\beta\right| > \sigma_{H,bound}$ overwhelm the coupling defined for the average structure, $\left|\beta\right|$. This is observed when no potassium is added; valinomycin is disordered such as to nearly completely randomize the orientation and distance between amide and ester groups, and $\left|\beta\right| < \sigma_{H,unbound}$. All of the carbonyl vibrations shift by a significant amount ($\Delta\epsilon_{bound} > \gamma,\sigma_{H,unbound},\sigma_{H,bound}$) due to the electrostatic field of the $\mbox{K}^+$ ion, which is notable for the amides because they are not directly involved in coordination. Analogous arguments for increased symmetry due to strong ion complexation has been found using Raman spectroscopy on the carbonyl groups in neat acetone and acetophenone.\cite{Giorgini2008,Giorgini2010}
The picture for $\mbox{K}^+$-binding in nonactin is similar. There is an $\mbox{K}^+$-bound state that contains a moderate amount of disorder ($\sigma_{H,bound} < \left|\beta\right|,\gamma$), which becomes much more disordered ($\sigma_{H,unbound} > \left|\beta\right|,\gamma$) and shifts drastically ($\Delta\epsilon_{bound} > \gamma,\sigma_{H,unbound},\sigma_{H,bound}$) upon loss of $\mbox{K}^+$. One difference however, is that in the absence of added $\mbox{K}^+$, nonactin retained some residual $\mbox{K}^+$, while valinomycin did not. Due to the strong binding of nonactin to $\mbox{K}^+$, it is difficult to completely remove $\mbox{K}^+$ from the solution and thus it is nearly always a two-component mixture. A fit of our FTIR spectra yields an equilibrium constant for nonactin-$\mbox{K}^+$ binding, $K_D$, of 3.25 +/- 0.75 mM in deuterated methanol, which can be compared to a literature value of 97 $\mu$M in protonated methanol.\cite{Izzat1985} This number can be greatly overestimated due to residual bound potassium, because the concentration of nonactin in these experiments was on the order of $K_D$. Further experiments on fully desalted nonactin would be required to fully disentangle the concentration-dependent spectra.
Although the calculations capture the salient features of the bound and unbound FT IR, Raman, and 2D IR spectra, there remain some disagreements. Figure~\ref{spectroscopy_model_compounds}a, b show that the nonactin spectrum undergoes broadening to the red, while in the corresponding calculations (Fig.~\ref{spectroscopy_model_compounds}i, j), all of the changes resulting from ion binding are to the blue of the unbound peak. This agreement can be improved by assuming that the no-added-$\mbox{K}^+$ ensemble is a two-component mixture of unbound and residual $\mbox{K}^+$-bound nonactin, but where the $\mbox{K}^+$-bound nonactin is more disordered than the same species at high $\mbox{K}^+$ concentration.
\section{Modeling vibrational spectroscopy of the KcsA selectivity filter}
In order to be able to extend the results on spectroscopic binding signatures of the model compounds onto the selectivity filter and to predict the expected vibrational spectra with and without isotope labels computational modeling should accompany the spectroscopic studies. We now apply our existing atomistic structure-based framework to model the IR and 2D IR spectra of KcsA in three different coordination states. The obtained changes in intensities, linewidths, frequencies, and cross-peaks can be compared to experimental data to confirm or reject the proposed structure. The amino acids of interest in the selectivity filter comprise 20/408 of the protein, and signatures of switching between coordination sites would require resolving changes to these vibrations. Such models are also indispensable for the selection of isotope labels as they can predict how isotope-labeled oscillators will shift as a result of the electrostatic contributions of $\mbox{K}^+$ binding. In this section, we model these features of the experiment computationally using molecular dynamics simulations of KcsA, and calculations of the amide I FT IR and 2D IR spectra.
\begin{figure}[t]
\centering
\begin{center}
\includegraphics[width=11cm]{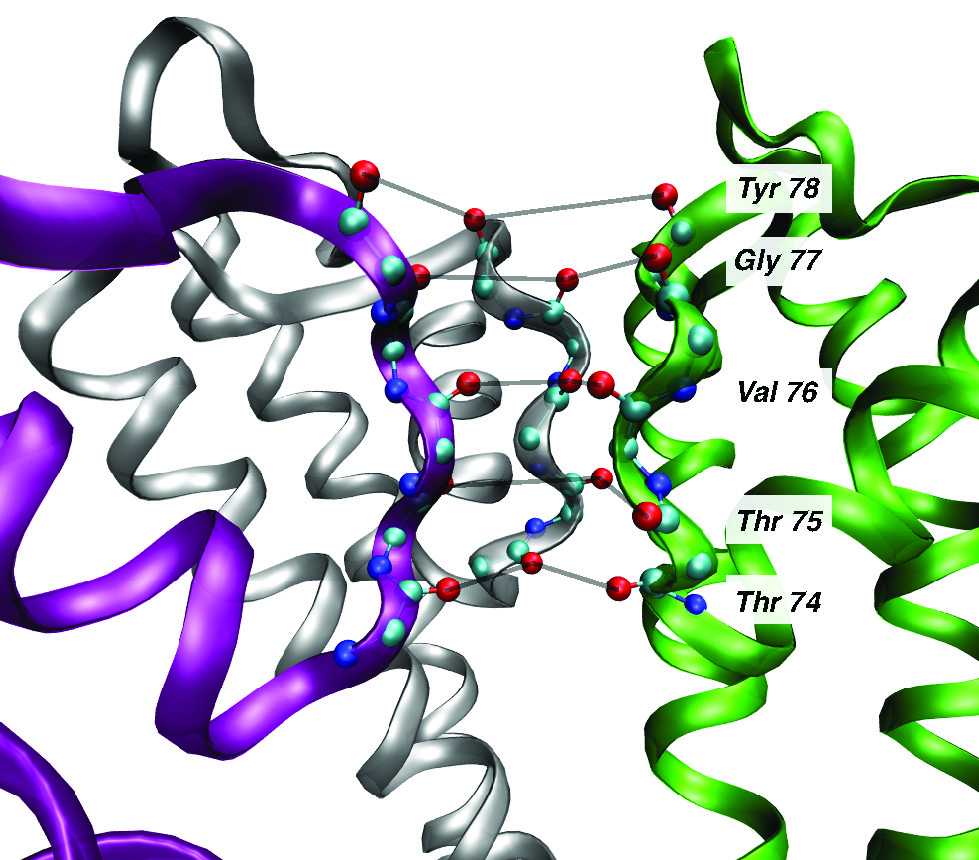}
\caption{Atomic structure of the selectivity filter backbone residues (Chains A-C are pictured, and chain D was removed for visual clarity). Lines are drawn between the carbonyl group oxygen atoms of the homologous residues in different chains. There are five rings of carbonyl oxygen atoms, each ring contains one carbonyl atom from each chain. The four gaps in between these five rings of oxygen atoms define the $\mbox{K}^+$ and H$_2$O coordination sites. Each of the pictured carbonyl residues is a candidate for an isotope-labeling strategy to highlight the difference between the KWKW and WKWK ligand states.}
\label{selectivity_filter}
\end{center}
\end{figure}
Figure \ref{selectivity_filter} shows the five selectivity filter residues responsible for the $\mbox{K}^+$ and H$_2$O binding. The coordination sites are roughly located between the rings of oxygen atoms. We will define the state KcsA-WKWK as having water in the site between Thr74/Thr75, potassium between Thr76/Val76, water between Val76/Gly77, and potassium between Gly77/Tyr78 with Thr74 being closer to the intracellular side and Tyr78 closer to the extracellular side. Along with the similarly defined KcsA-KWKW and KcsA-WWWW ligand states, these were the three states simulated for the KcsA channel. KcsA-KWKW and -WKWK have been postulated to be important stable states for the throughput of $\mbox{K}^+$ ions (defined as the 1,3 and 2,4 states, respectively, in Ref.\cite{Morais-Cabral2001}). KcsA-WWWW was simulated as a spectroscopic background state, which may also be useful in interpreting control experiments.
\subsection{Simulating FT IR and 2D IR Spectra using Molecular Dynamics Simulations}
\small
Two-dimensional infrared spectra of the KcsA potassium channel in various ion coordination states and with various isotope-labeling strategies were calculated using the low $\mbox{K}^+$-binding crystal structure\cite{Zhou2001} using GROMACS 3.3.1,\cite{Lindahl:2001p1008,Berendsen:1995p1039} and the OPLS/AA force field.\cite{opls} The antibody Fab fragment (A and B chains) and crystallization co-factors were deleted. Since the crystal structure was missing atoms, the WHATIF web server was used to fill in missing residues.\cite{Vriend1990} The structure was then protonated at pH 7, and solvated with SPC/E water\cite{BERENDSEN:1987p1279} in a 10 nm cubic box with periodic boundary conditions. Simulations were run in the NPT ensemble at 1 atm and 300 K. All bond lengths were constrained and the simulation time step was 2 fs. Three ligand sets for the four binding sites of KcsA selectivity filter were considered. The ligand sets (hereafter referred to as WKWK, KWKW, and WWWW) were constructed using the same protein crystal structure by placing the $\mbox{K}^+$ ion or oxygen of the water in the center of the eight coordinating oxygen atoms. 8-12 $\mbox{Cl}^-$ counterions were added for charge neutrality. Each system was then allowed to energy-minimize for 1000 steps and unconstrained dynamics were run for 10 ns. While KcsA is a membrane protein, it was simulated in water for computational simplicity because the goal was only to sample short-time fluctuations about the crystal structure. No significant deviations from the crystral structure occurred during the simulation. It was observed that the water in KcsA-WWWW and -KWKW escaped the ion channel, so an external, 10 kJ mol$^{-1}$nm$^{-2}$, weak harmonic restoring force was added constrain each of the oxygen atoms of the water ligands to their coordination site in the crystal structure in the z-axis only (coinciding with the conduction axis of KcsA). To put this into context, such a force constant gives rise to a 0.05 kJ/mol energy penalty for a 1$\AA$ translocation out of the filter; the native potential of $\sim $60 mV across a $\sim$5 nm cell membrane yields $\sim $0.12 kJ/mol for the same motion of a particle with 1\textit{e} charge. This energy is $\sim $0.17 kJ/mol when one considers the time-dependent action potential change of $\sim $90mV that occurs during the biological function of neurons, which suggests that the applied constraints fall within the energy scale of existing external perturbations. These constraints were sufficient to gather structures for 10 ns, maintaining the coordination state all throughout.

\small
Full Cartesian coordinates of the entire simulation box were saved 1/50 fs, and 12,500 of these frames were used as inputs to calculate a local amide Hamiltonian using the methods described in Ref~\cite{Ganim2010}. In this model, each amide I oscillator in the backbone generates a vibrational site. The intrinsic frequency of each site is determined by the electrostatic environment that it senses. Coupling between each vibrational site has a through-bond component and a through-space component; all aspects of the local amide Hamiltonian are parameterized by quantum mechanical calculations.\cite{Jansen:2006p899,Bour:2003p902} This type of modeling has been used to calculate FT IR and 2D IR spectra of peptides and proteins, and to interpret experiments.\cite{Ganim2008,Choi2007, Liang2011}

\small
One distinguishing feature of this work is the size of the system studied; it is the largest protein to have its 2D IR spectrum calculated. The full KcsA system comprises 412 residues, and therefore 408 amide I oscillators. The two-quantum Hamiltonian for this system is of dimension 83,028 and would require 26 Gb in memory to store while diagonalizing. Therefore, we chose to approximate the system by block diagonalizing the Hamiltonian; one block was defined using residues 32 to 63 (including the selectivity filter of KcsA, the short $\alpha$-helix preceding it, and the random coil regions flanking them) in all four chains, for a total of 128 residues. The rest of the protein (280 residues) was block diagonalized as previously described,\cite{Ganim2010} using a cut-off of 4.0 $cm^{-1}$. The FT IR spectra could be calculated without block diagonalization. An 18 $cm^{-1}$ static redshift is applied to all the spectra to compare more favorably against experimental amide I spectra, which arises from a systematic error in the quantum calculations.\cite{Bour:2003p902,Jansen:2006p899,Schmidt2004} For isotope labeling calculations, the selected oscillators were given a 65 $cm^{-1}$ redshift, corresponding to the effect of a $^{13}$C-$^{18}$O isotope label, and the appropriate block of the Hamiltonian was rediagonalized.

\small
A consistency check for these simulation constraints is the similarities in the FT IR and 2D IR spectra; WKWK has no constrained ligands, KWKW has two constrained ligands, and WWWW has four. Due to the fact that the FT IR and especially 2D IR spectra are sensitive to minor changes, the close comparison among these spectra shows that the effect of changing ligands is a minor perturbation. More importantly, artifacts arising from the different restraint conditions on KWKW and WKWK would appear in the calculated difference spectra (WKWK-WWWW and KWKW-WWW in Figures~\ref{kcsa_ftirs}b and \ref{kcsa_2D IRs}), which were found to be nearly identical. Nonetheless, a more accurate, restraint-free calculation in a lipid environment would provide direct evidence, and would further predict ion translocation kinetics, mechanism, and those associated spectral signatures. As a check for the spectral approximations, one can note the close consistency between the features observed in the FT IR spectra (calculated without block diagonalization approximations) and 2D IR spectra (whose Hamiltonian was isolated into several blocks). Moreover, it was observed that the majority of the changes arise from the 128 residue block containing the selectivity filter and ligands, which were maintained as one block of the Hamiltonian in all calculations.
\normalsize
\subsection{FT IR and 2D IR Spectra of KcsA-WKWK, KWKW, and WWWW}
\begin{figure}[t]
\centering
\begin{center}
\includegraphics[width=7 cm]{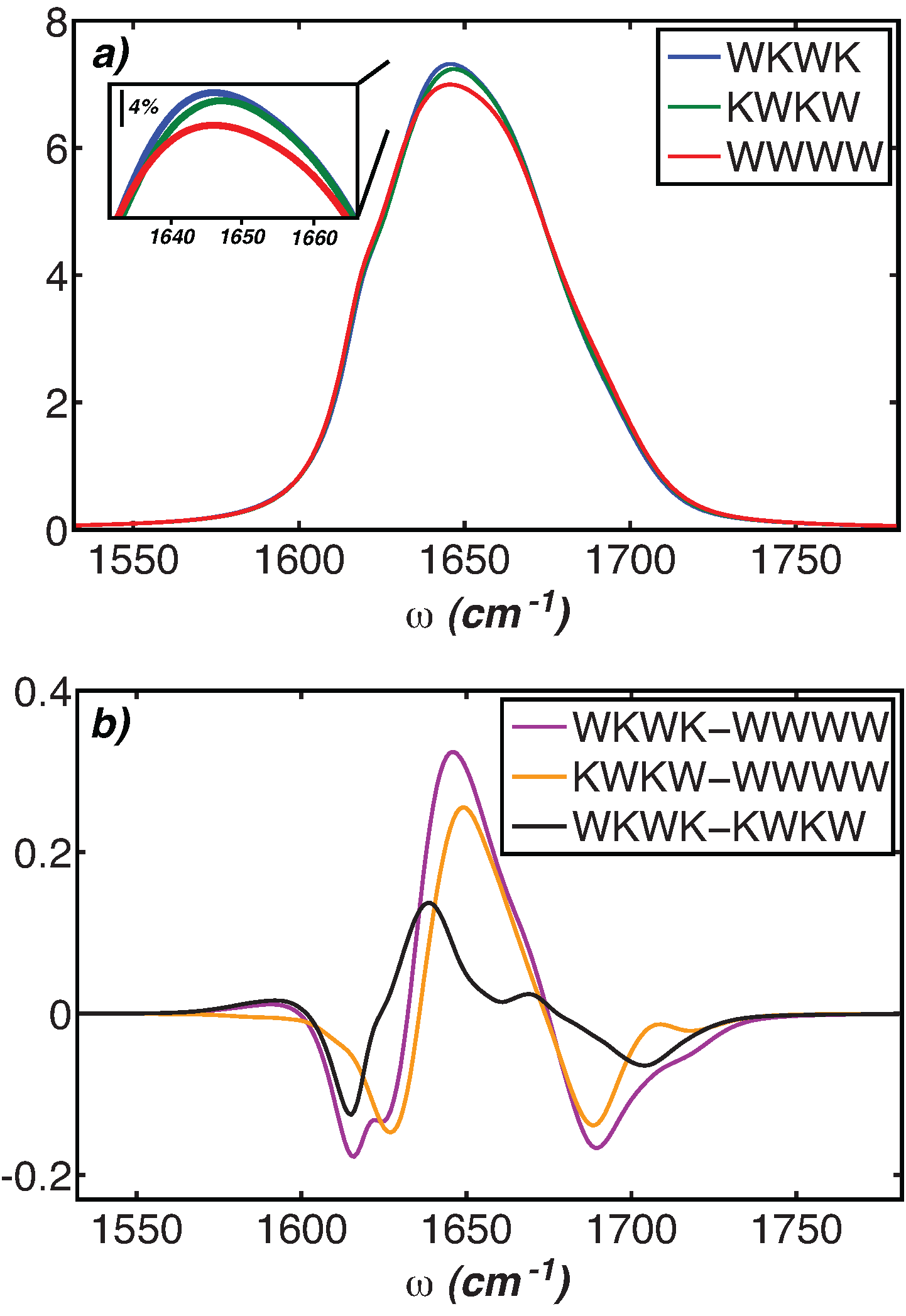}
\caption{FT IR spectra of KcsA-WKWK, KWKW, and WWWW (a) and difference spectra (b). Inset in (a) shows an expansion of the region with largest changes. These spectra demonstrate the modest changes induced by potassium translocation in the absence of isotope labeling.}
\label{kcsa_ftirs}
\end{center}
\end{figure}

\begin{figure}[t]
\centering
\begin{center}
\includegraphics[width=14 cm]{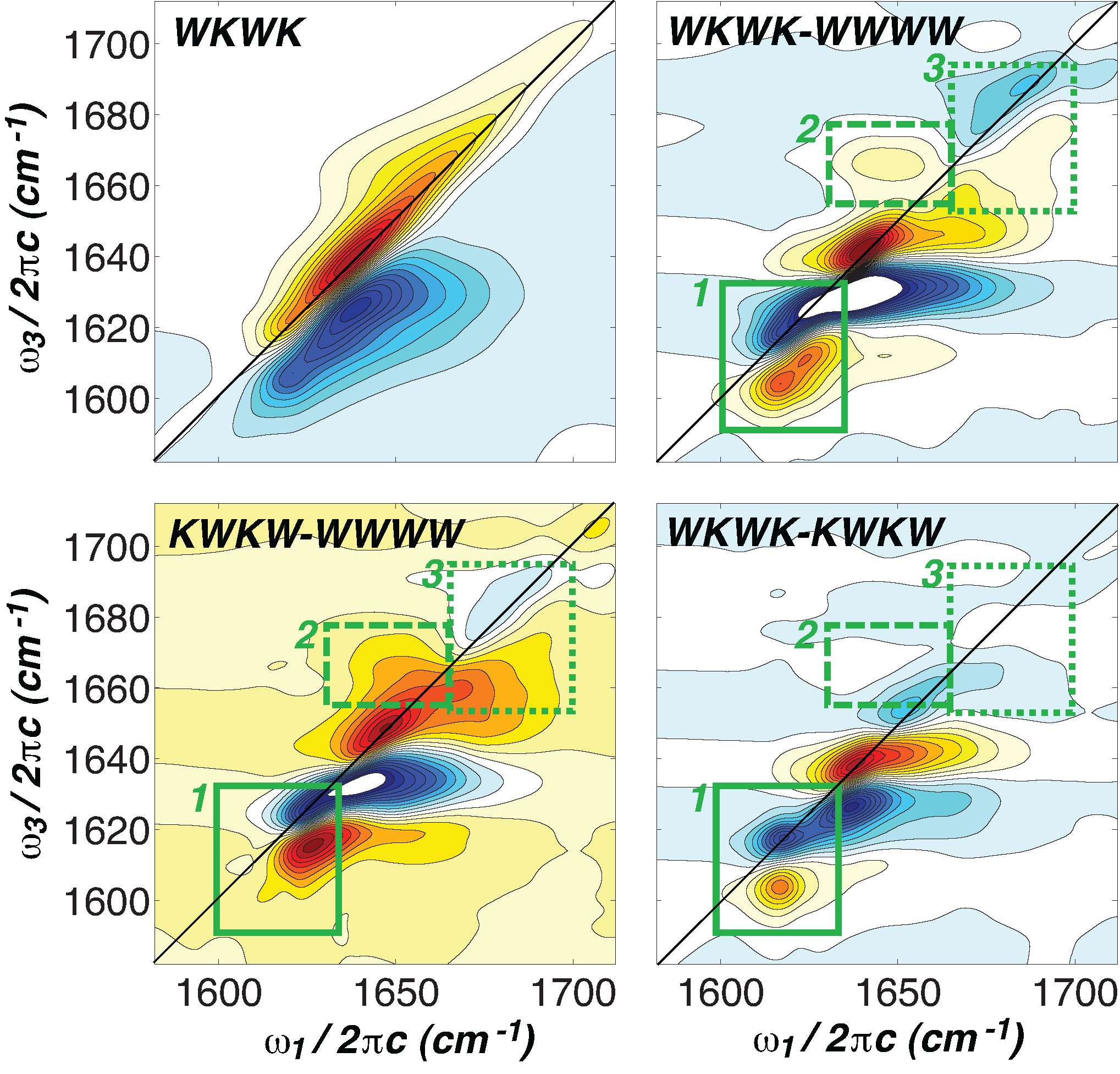}
\caption{2D IR spectra of KcsA-WKWK, KWKW, and WWWW and difference spectra. Green boxes highlight spectral regions for comparisons, which are defined as box 1 (solid, $\omega_1=1592 - 1632\ cm^{-1}$, $\omega_3=1600-1636\ cm^{-1}$), box 2 (dashed, $\omega_1=1630 - 1666\ cm^{-1}$, $\omega_3=1655-1677\ cm^{-1}$), and box 3 (dotted, $\omega_1=1664 - 1700\ cm^{-1}$, $\omega_3=1651-1695\ cm^{-1}$). Contours plotted from the minimum to the maximum of the WKWK signal in 8.3\% increments for WKWK and at 10\% of those levels for all difference spectra. All spectra calculated for a waiting time of 0 fs. These spectra demonstrate the frequency regions giving the biggest changes between the different ligand states.}
\label{kcsa_2D IRs}
\end{center}
\end{figure}

Based on the number of amide I carbonyls coordinating ligand atoms (20) compared to the total number of amide I oscillators (408), one would expect that a substantial change in the frequency or fluctuations of every oscillator should yield an overall 4.9\% change in the signal. Our understanding of the model compounds suggest that upon $\mbox{K}^+$ binding, the carbonyl residues should form a more ordered ensemble. In Figure~\ref{kcsa_ftirs}a, the FT IR spectra of each of the three simulated states are plotted. The spectra appear very similar, with differences appearing to be changes in intensity on the 2-5\% level. Since the same number of amide I oscillators are included in each calculation and the magnitude of each oscillator was not variable, the spectra must be frequency-integrable to the same value. Figure~\ref{kcsa_ftirs}b demonstrates this by showing that the apparent intensity changes are due to changes in broadening. Relative to the water-filled channel, binding potassium in the KWKW or WKWK state leads to a narrowing of the line, evidenced by an increase at $\approx1650\ cm^{-1}$, and decreases to the red, at 1615-1625 $cm^{-1}$, and to the blue at 1688 $cm^{-1}$. The spectra of the two potassium bound states seem to indicate that potassium is stronger-bound in WKWK than KWKW; in part a, the intensity of WKWK is higher than KWW and in part b, the difference features (WKWK-KWKW) are similar to those relative to no potassium (WKWK-WWWW). 
Similar to the results from the FT IR spectra in Figure~\ref{kcsa_ftirs}, the differences among the 2D IR spectra of KcsA-WKWK, -KWKW, and -WWWW are difficult to detect by visual inspection; Figure~\ref{kcsa_2D IRs} shows a 2D IR spectrum of KcsA-WKWK that is representative of these three spectra. The subtle features that distinguish WKWK, KWKW, and WWW mostly appear in the amount of diagonal elongation, which indicates that the differences are dominated by changes in structural disorder. The FT IR spectra showed that adding potassium caused a line-narrowing effect- an intensity increase at the peak and loss of intensity from the wings of the spectrum. This loss-gain-loss is a clear signature of line narrowing, which also show in the 2D IR spectra; each 2D IR spectrum shows a negative doublet in the low frequency region of box 1, a gain feature in the center of the spectrum at 1650 $cm^{-1}$, and relative to the water-only channel, WKWK and KWKW also show a loss in the higher frequency region of box 3. The feature in box 1 can be taken as a sign of stronger potassium binding, and is consistent with the low frequency loss feature seen in the FT IR spectra. In box 3, there is evidence for a broader, high-frequency loss feature that is roughly similar in KWKW and WKWK, and therefore dependent on potassium and water-bound to the selectivity filter, but not the sequence. Similarly, there is an off-diagonal gain feature in the dashed box in KWKW and WKWK relative to water, which is a sign of the line narrowing upon potassium binding, but also doesn't help distinguish WKWK and KWKW.
\subsection{Distinguishing WKWK and KWKW in FT IR and 2D IR Spectra with Isotope Labels}
To enhance the spectral differences between KcsA-WKWK and KcsA-KWKW, different isotope labeling combinations were simulated. The general idea is that since the amide I oscillators that are directly coordinating $\mbox{K}^+$ are subject to strong electric field changes as ions are throughput, perhaps including isotope labels would allow for background-free observation of the frequency shifts these oscillators undergo. Since there are five amino acids responsible for ion coordination in the selectivity filter, we simulated the FT IR spectra for all ($2^5=32$) possible isotope-labeling combinations of these oscillators. It was assumed that incorporating an isotope label would lead to uniform enhancement at this site in all chains. The nomenclature in the subsequent discussion describes the location(s) of a $^{13}$C-$^{18}$O isotope label (65 $cm^{-1}$ redshift) among the five positions in the selectivity filter: Thr74, Thr75, Val76, Gly77, Tyr78. The state without isotope labels is called 00000; one isotope label at Thr75 is 01000; three isotope labels at Thr74, Gly77, and Tyr78 is 10011. Depending on the synthetic strategy, only a subset of these isotope labeling combinations may be feasible, a discussion for which appears in the Outlook section.
After all of the isotope-labeled FT IR spectra were calculated, the differences WKWK-KWKW were calculated and the norm of the difference spectra between 1550 - 1610 $cm^{-1}$, the expected region for background-free, isotope-labeled peaks,
\begin{eqnarray}
\left(\int_{1550\ cm^{-1}}^{1610\ cm^{-1}} \left| I^{WKWK}(\omega)-I^{KWKW}(\omega) \right|^2\right)^{1/2}. \label{normdiff}&&
\end{eqnarray}
\begin{figure}[t]
\centering
\begin{center}
\includegraphics[width=16 cm]{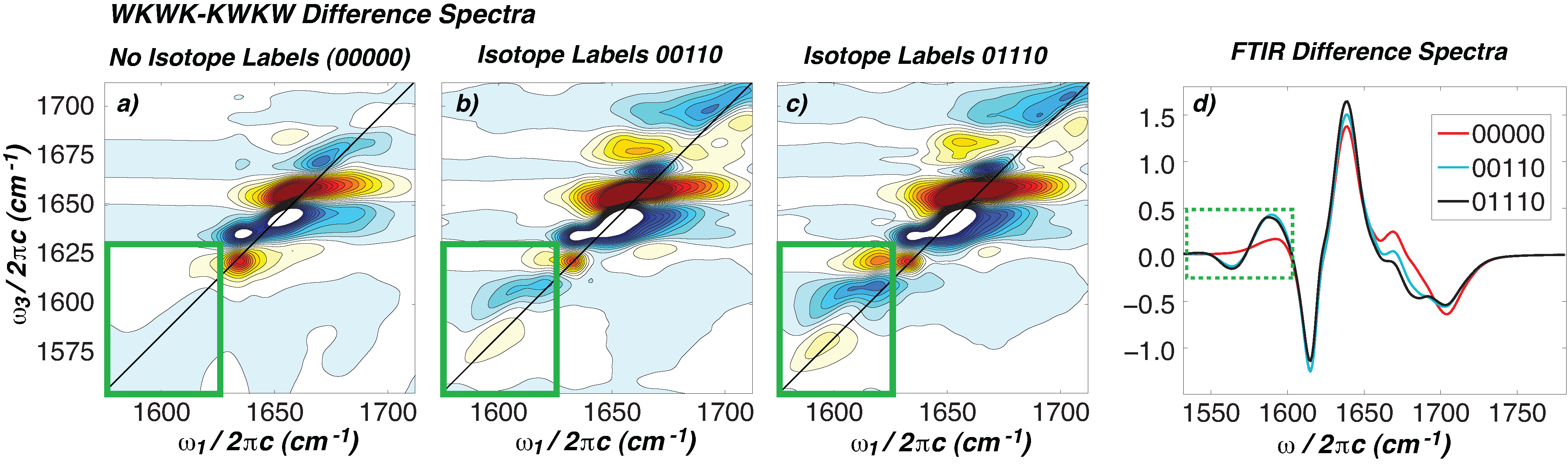}
\caption{2D IR and FT IR spectra of WKWK-KWKW for different isotope labeling combinations. The solid green boxes ($\omega_1=1575 - 1625\ cm^{-1}$, $\omega_3=1555 - 1633\ cm^{-1}$) highlight the additional, isotope-labeling-induced peaks, which can be used to discriminate between the KWKW and WKWK states. Contours plotted from -6\% to 6\% of maximum of the WKWK signal, in 0.05\% increments. In the FT IR spectra, the dashed box shows the region that was used to choose the most effective isotope labeling combinations. All spectra calculated for a waiting time of 0 fs.}
\label{isotope_spectra}
\end{center}
\end{figure}
This region of the spectrum is highlighted in Figure~\ref{isotope_spectra}d. On the basis of maximizing the quantity in Equation~\ref{normdiff}, two isotope labels were chosen to have the corresponding 2D IR spectra calculated. Note that with the aforementioned block-diagonalization strategy, only the contribution from the smaller, 128 oscillator block containing the selectivity filter needs to be recalculated. The best isotope labeling combinations for enhancing the difference between KcsA-WKWK and KWKW were found to be Val76-Gly77 (00110) and Thr75-Val76-Gly-77 (01110). The spectra from these isotope labels appear in Figure~\ref{isotope_spectra}.
The isotope labels cause changes to the WKWK-KWKW difference spectrum in several ways (The difference spectrum without isotope labels is shown in Figure~\ref{isotope_spectra}a for comparison). The most distinct change is the appearance of two features in the far red region of the spectrum, which is highlighted with a green box in Figure~\ref{isotope_spectra}a-c. Two doublets appear at $\omega_1$=1600 $cm^{-1}$ and $\omega_1$=1625 $cm^{-1}$, where the lower frequency one is oppositely-signed. This features are consistent with the loss-gain feature in this region of the FT IR spectrum (Figure~\ref{isotope_spectra}d). These changes, which directly result from isotope-shifted oscillations, are on a similar magnitude level as a loss feature on the far blue side of the spectrum ($\omega_1$=1700\ $cm^{-1}$) and some additional off-diagonal changes in the middle of the band. Because there is no significant change to the distances and orientations between carbonyl groups from the WKWK to KWKW structures, this spectral change is attributed purely to the change in electrostatic environment as potassium and water switch coordination sites rather than due to a change of coupling. It is notable that the isotope-induced changes are smaller than the naturally observed changes due to switching of the ligands from WKWK to KWKW; however, isotope-labeling induces changes in a background-free region of the spectrum at $\omega_1$=1600\ $cm^{-1}$, which can lead to unambiguous assignment to the isotope-labeled oscillators.
It is noteworthy that the 30 other isotope-labeling combinations yielded even smaller changes than those shown in Figure~\ref{isotope_spectra}. 27 of the calculated spectra yielded almost no change in the low-frequency region under inspection here. For an oscillator that has an intrinsic absorption on the blue side of the band, even a 65 $cm^{-1}$ redshift is not sufficient to isolate it from the remainder of the amide I band.

Our main finding from simulations of the 2DIR spectra of the isotope labeled selectivity filter is the identification of distinct synthetic targets, the 00110 and the 01110 combinations, that lead to spectroscopic features in the background free region of the amide I spectrum (Figure~\ref{isotope_spectra}). These isotope-labeled ion channels allow for the effect of translocating $\mbox{K}^+$ ions in transient 2DIR experiments to be probed via distinct amide I spectroscopic features. 2D IR experiments can be used to characterize the excitonic nature of carbonyl resonances. In a manner analogous to 2D electronic experiments on the FMO photosynthetic complex, a combination of cross-peaks and splitting patterns for the C=O vibrations will provide a characterization of how synchronized the vibrational motions along the selectivity filter are. Also, waiting time transient 2D experiments, in which the time between excitation and detection is varied, will allow for the presence of coherence to be deduced via the beating of cross peaks between those modes in the background free region shown in Figure~\ref{isotope_spectra}.

In that regard the simulation results presented in this work confirm the feasibility of a critical step towards experiments that could ultimately evaluate the presence and relevance of quantum coherence for the function of the selectivity filter. Since it is quite labor-intensive to incorporate such isotope labels in proteins, especially a membrane protein such as this, it is worthwhile the use the most accurately available calculations to help design the experiment.

\subsection{Conclusions from KcsA Modeling and Comparison to Model Compounds}
One of the most prominent spectral changes to nonactin and valinomycin upon adding potassium is the line narrowing due to the increased rigidity afforded by ion coordination. In the FT IR and 2D IR spectra of the KcsA filter, a similar narrowing is seen when the ion channel contains potassium relative to water only. Amongst the fluctuations in the other amide oscillators in the protein, this remains a minor effect, on the 4\% level. For the 2D IR spectra, these intensity changes are on the 10\% level, and may allow one to experimentally distinguish potassium binding.
Concomitant with the increased rigidity due to potassium binding in the model compounds was an electrostatic shift to frequencies of oscillators coordinating potassium. While this shift cannot be detected in the native KcsA protein, isotope labeling combinations have been identified that highlight these changes, and produce difference signals that can be used to identify KWKW to WKWK switching. It was concluded that this change to the vibrational spectrum results primarily from the change in electrostatics; local changes, such as rotations of the carbonyl groups (which are included in our simulations and were previously observed\cite{Gwan2007}) contribute less to the changes in the vibrational spectrum than the different electrostatic environment caused by migration of the $\mbox{K}^+$ ion.
\subsection{Outlook}
Future experiments on the model compounds will allow for the controlled testing of a number of complications that are expected in experiments on the ion-channel. Mixtures of model compounds with proteins and lipids can be used to quantify the sensitivity limits of the experiment, test which isotope labeling scheme will provide the most information, and generally optimize experimental conditions for the protein sample. Furthermore, the model systems will be the point for testing the efficiency, practicality and the time scales for photo-release compounds to be employed in transient 2D IR studies.
\vspace{0.25cm}
There is also a second class of model compounds, 4xPPs, which are models for the selectivity filter featuring 4-fold symmetric TVGYG polypeptides built on a p-tert-butyl-calix[4]arene scaffolding \cite{Rivas2001}. This recreates a channel which binds $\mbox{Na}^+$ and $\mbox{K}^+$. These models can be synthesized with isotope labeling patterns and hence allow us to test sensitivity to ion binding location and motif. Isotope-labeled 4x(TVG**YG*) and similar isotopomers will be synthesized to probe and quantify exciton formation at the labeled region of the filter to determine $\mbox{K}^+$ binding patterns in concentrated salt solutions (where * refers to $^{18}\mbox{O}$ and ** refers to $^{13}\mbox{C}^{18}\mbox{O}$). For instance, if site 1 is bound to $\mbox{K}^+$, is site 3 also bound? Does one only see the 1,3 (GG) and 2,4 (YV) binding motifs as predicted by crystallography results, or are singly bound species present? Such studies can form the first layer of information to the equilibrium binding structures within the selectivity filter.
\vspace{0.25cm}
The calculations in this manuscript outline the changes that can be expected in the KcsA channel upon potassium binding, and switching between the WKWK and KWKW states. They also suggest a set of isotope-labeled proteins, which can be synthesized to further resolve the difference between the WKWK and KWKW state. As the next step one could test whether that the signatures of ion-binding observed in the model compounds and calculations are seen in KcsA. For this purpose 2D IR spectroscopy is unique because of its structural resolution, phase-sensitivity, and picosecond time-resolution, and phase-sensitivity. To study the structure and dynamics of the ion channel as potassium is transmitted in real time, it would be necessary to abruptly introduce potassium to the system, and subsequently record vibrational spectra on timescales faster than the ion transport rate. To this end, caged potassium can be used as an excitation mechanism, coupled with 2D IR spectra recorded as snapshots.\cite{Chung2009} More information can be obtained by uncaging potassium during the waiting time to watch vibrational modes evolve as potassium displaces water in the channel through the formation of cross-peaks.\cite{Bredenbeck2003} Once a steady state of potassium throughput is achieved, it is expected that the phase and time-dependence of the cross-peaks would allow one to distinguish between kinetic hopping between KWKW/WKWK or the formation of a quantum coherent state.
To translate these results into a synthetic strategy, one would ideally be able to site-specifically label any desired combination of residues. In principle, such a strategy could proceed by introducing the isotopes to small regions of the protein using solid-phase peptide synthesis, and ligating them to produce the desired product.\cite{Dawson1994} A far more feasible strategy is to use a selective labeling approach in which all amino acids of a certain type in the protein are labeled by introducing the isotope labeled amino acids as a precursor during protein expression. For example, the isotope-labeling combination predicted to produce the largest change, 00110, could be produced by labeling all valine and glycine residues. A similarly large change was observed for the 01110 isotope-labeling combination, but this would require labeling of Thr75 and not Thr74. This strategy would necessarily produce isotope-labeled sites in the protein outside the selectivity filter. It is expected that since these sites would not change during $\mbox{K}^+$ translocation, they would contribute to the existing background in the 1550-1610 $cm^{-1}$ region and not influence the KWKW-WKWK difference spectra. Similarly, the natural abundance of $^{13}C$ would also contribute to the background signal. Because both of these strategies can be pursued with nearly 100\% efficiency, there is no need to consider heterodimers between labeled and unlabeled monomers.
But even if the coherences are found, probably the most intriguing aspect of quantum coherence in biological system is to understand their relevance for the biological function. First, it is conceivable that quantum coherence can be found in many biological systems but only as an inevitable result of the involved spatial and temporal scales without being exploited for the biological function. In that sense quantum coherence could turn out to be only 
epiphenomena. But even that should turn out not to be the case, it would be fascinating to see if quantum coherence has been exploited by nature for the same functional tasks across different systems or whether it is involved in diverse sets of functional mechanism. Are there any underlying principles that one can use to predict when and for which purpose a system might exhibit quantum effects? 
\vspace{0.25cm}
For quantum coherence to not only have a trivial role in biological function, i.e. based on the principle that ultimately all matter is built up elementary particles which exhibit well known quantum behaviour that is required for the stability of biomolecules and therefore for biology, the coherence must have a more direct or ``unexpected'' consequence on the function. This often implicates that the expected quantum coherence should be observed on a more macroscopic, but still nanometer scale, as it was also confirmed in the case of the photosynthetic complex. Coherences at such scale and subjected to environmental noise decohere quickly. Hence we expect in order to understand the functional relevance of the quantum coherence for the function it is will be also necessary to study the transition between the quantum to the classical regime in the dynamics of macromolecular interactions. In fact we believe that the interplay between quantum coherence and the environmentally induced decoherence is how in many cases quantum effects are involved in a functional role in biological systems. Such interplays can ultimately lead to classical states that carry a ``signature of quantum coherence'' and that are conventionally known to mediate the biological function. 
\ack
This work was supported by Alexander von Humbolt Stiftung, a grant from the National Science Foundation (CHE-0911107), Howard Hughes Medical Institute and Wiener Wissenschaft und Technologie Fond (WWTF). 

\end{document}